\input{aipcheck}

\documentclass[final            
  ]
  {aipproc}

\layoutstyle{8x11double}

\begin{document}

\title{The Cepheid Distance Scale: Recent Progress in Fundamental Techniques}


\author{Thomas G. Barnes III}{
  address={The University of Texas at Austin, McDonald Observatory, 1 University Station, C1402, Austin, Texas, 78712-0259}
}

\begin{abstract}
This review examines progress on the Pop I, fundamental-mode Cepheid distance scale 
with emphasis on recent developments in geometric and quasi-geometric techniques for 
Cepheid distance determination.  Specifically I examine the surface 
brightness method, interferometric pulsation method, and trigonometric 
measurements.  The three techniques are found to be in excellent agreement for distance measures in the Galaxy. The velocity p-factor is of crucial importance in the first two of 
these methods.  A comparison of recent determinations of the $p$-factor for Cepheids 
demonstrates that observational measures of $p$ and theoretical predictions agree within their uncertainties for Galactic Cepheids.  
\end{abstract}

\maketitle


\section{Introduction}
  In the near-century since Henrietta Leavitt's announcement of the Cepheid period-luminosity relation (Leavitt \& Pickering 1912), enormous progress has been made in our understanding of the observational properties and physical origin of Cepheid pulsation.  A delightful and thorough presentation of the early history of what is now being called  the Leavitt Law is given by Fernie (1969). But one aspect of the relation has proved elusive -- a calibration of Cepheid luminosities based on fundamental geometry.   Due to the large distances to Cepheids -- other than the overtone pulsator Polaris near 130 pc, the closest is $\delta$ Cep itself at 273 pc -- distance determinations have depended, first upon statistical parallaxes, and later, upon the presence of Cepheids in galactic clusters.  Neither of these methods can be considered \emph{fundamental} in the geometric sense.  Recent developments have changed that situation.  
  
In this paper I will examine the techniques of fundamental distance measurement of Cepheids, compare the results from those techniques, and discuss the potential systematic error  common to two of the techniques.  I will restrict my discussion to fundamental-mode Cepheids only.

\section{Fundamental Techniques }
Three methods qualify as geometric or quasi-geometric determinations of Cepheid distances: surface brightness pulsation distances, interferometric pulsation distances, and trigonometric parallaxes.   I omit here discussion of the Cepheid distances determined by means of the maser in NGC 4258 as that result is better used as a check on the other three.

The distinction I make between geometric and quasi-geometric is the following.   Trigonometric parallaxes are based on geometry.  Some may quibble that the adjustment from relative to absolute parallax is not geometric, but I would argue that the distances to the reference stars may be traced back to trigonometric parallax calibration.  On the other hand, the quasi-geometric methods (comparison of linear diameters to angular diameters to determine the distance) would be geometric except for complications that are not geometric.  In the case of the surface brightness pulsation method, this is the $p$-factor that converts observed radial velocity into stellar pulsation velocity.  In the case of interferometric pulsation distances there is the limb darkening correction to the uniform-disk angular diameter as well as the $p$-factor.

\subsection{Surface Brightness Distances}
The surface brightness technique is an extension of the work by Baade (1926) and by Wesselink (1946, 1969).  The first two of these papers established a practical method for determining the mean radius of a Cepheid without knowing the actual surface brightness.   Phases of equal color index are assumed to be phases of equal surface brightness.  The difference in magnitude between the two phases is then dependent only on the ratio of the radii at the two phases.  The difference  in radii at the two phases may be determined by integration of the radial velocity curve, appropriately converted to a pulsational velocity  curve. Application to multiple phases yield the mean radius.  This is the classic Baade-Wesselink method. 

Wesselink (1969) later determined actual surface brightnesses for eighteen stars of measured angular diameter.  Although no Cepheid  variables were among the sample, he assumed (with full acknowledgment of the risk) that the correlation between these surface brightnesses and $(B-V)$ would  apply to Cepheids.  From the mean $(B-V)$  of a Cepheid and his correlation he obtained the mean surface brightness; from the Baade-Wesselink method he obtained the mean stellar radius; a combination of these two provided the mean absolute magnitude.  This method depends upon a reliable determination of the color excess in order to infer the Cepheid surface brightness correctly from $(B-V)$ and also to compute the distance. Later studies found that the slope of the surface brightness -- $(B-V)$ relation determined from angular diameter stars did not agree with the slope for Cepheid variables (see for example, Thompson 1975).  That notwithstanding, Wesselink obtained a  distance to  $\delta$ Cep of 270 pc, in remarkable agreement with the recent trigonometric parallax distance of 273 pc.

Barnes \emph{et al.} (1976, 1977) extended the Wesselink (1969) approach in two ways.  Using a much  larger sample of measured angular diameters, they showed that the visual surface brightness $F_v$ correlated with $(V-R)$ much better than with $(B-V)$.  The tight correlation included stars of all luminosity classes, unlike the separation of supergiants exhibited in the $(B-V)$ correlation, and thus seemed likely to be applicable to supergiant Cepheid variables.   Secondly, their mathematical approach to the problem solves for the distance and radius simultaneously, unlike Wesselink's separate solutions. Not only is this approach more appropriate mathematically, but it also, when used with $(V-R)$, renders a distance that is essentially independent of the reddening. A one magnitude \emph{error} in the adopted interstellar extinction $A_v$ causes a 4\% error in distance. 

Significant improvement in the surface brightness method was introduced by Welch (1994).  He demonstrated that use of the infrared color index $(V-K)$ preserves the advantages of the surface brightness method (quasi-geometric, insensitivity to reddening) while reducing the uncertainties substantially.  Fouqu\'e and Gieren (1997) compared the $V,(V-R)$; $V,(V-K)$; and $K,(J-K)$ versions of the method and found that the three choices  yield very similar distances and radii but that the $V,(V-K)$ combination produces percentage uncertainties nearly an order of magnitude smaller than those for $V,(V-R)$.  As a result most researchers have adopted the infrared surface brightness method.   One problem with this choice is that the $V$ and $K$ photometric data are seldom acquired simultaneously, hence some interpolation scheme based on accurate knowledge of the period is needed to compute the colors.  
 
\begin{figure}
 \includegraphics[height=.3\textheight]{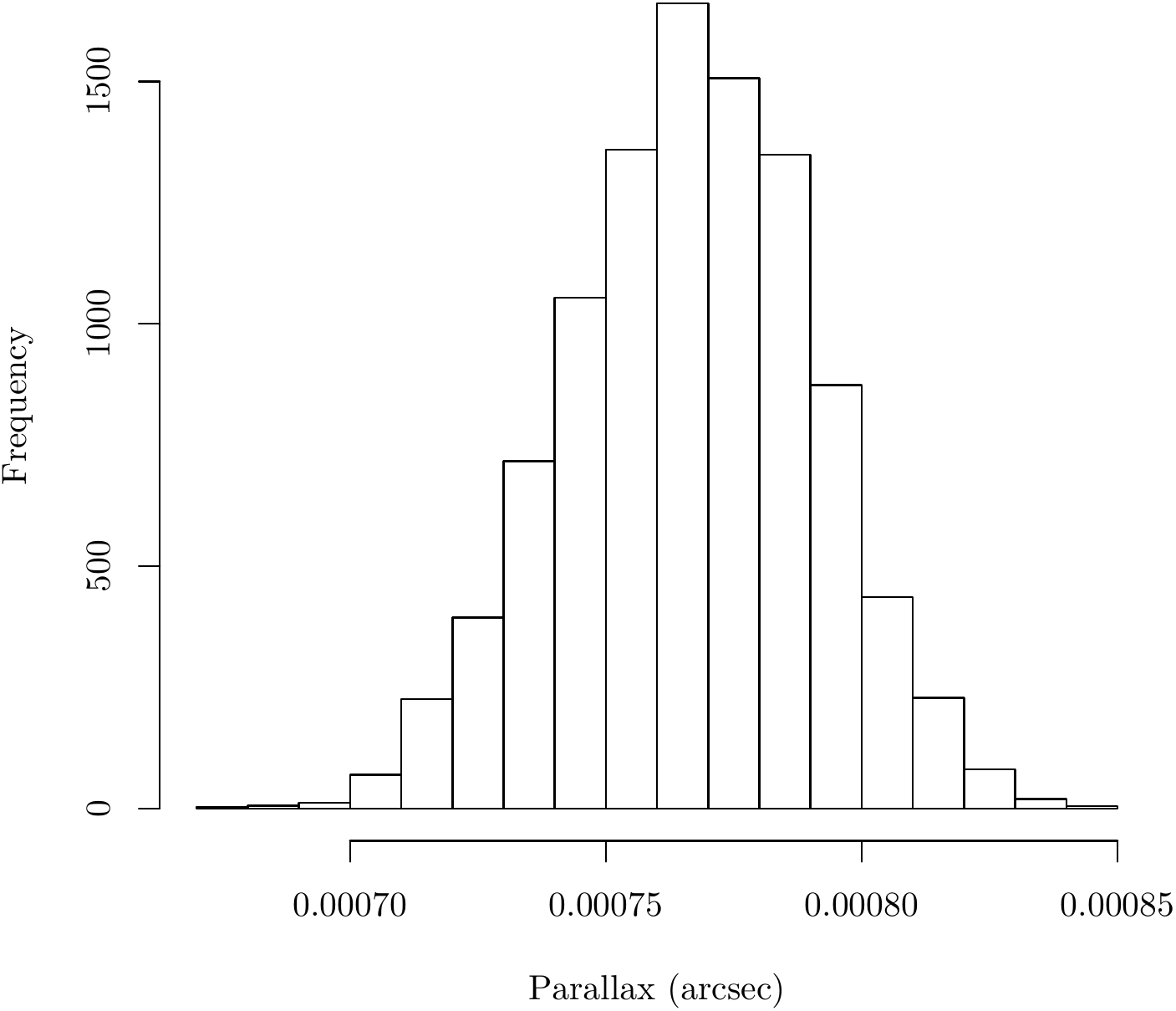}
\caption{Posterior marginal distribution of the parallax of T Mon. Figure from Barnes \emph{et al.} (2003). \label{fig1}}
\end{figure}

Despite  the shift to the infrared, the surface brightness method still suffered from two major problems.  None of the mathematical solutions by various researchers to the equations for determination of distance and radius from magnitude, color and radial velocity were rigorous,  and none of the angular diameters used to calibrate the surface brightness -- color relation were obtained from Cepheids.    These two issues were not fully addressed until very recently.   

A rigorous  and objective solution to the mathematics of the surface brightness equations was provided by Barnes \emph{et al.} (2003) using a Bayesian Markov-Chain Monte Carlo code.   This paper also provides a thorough discussion of the equations needed to solve for the distance and radius in the surface brightness method.  Their analysis objectively selects a model for the radial velocity curve, correctly propagates the radial velocity uncertainties through the analysis, correctly handles the problem of uncertainties in both the inferred angular diameters and the computed linear displacements, and averages over the probabilities associated with all the models for distance and radius.  The latter is demonstrated in Figure 1. The adopted parallax is the expectation value of the posterior marginal distribution and its uncertainty is given by the breadth of that distribution.   In a later paper Barnes \emph{et al.} (2005) showed that the linear-bisector solution to the surface brightness equations adopted by some researchers gives the same results for distance and radius as the Bayesian solution, but underestimates the uncertainties considerably and lacks the internal checks of the Bayesian approach.  

\begin{figure}
 \includegraphics[height=.3\textheight]{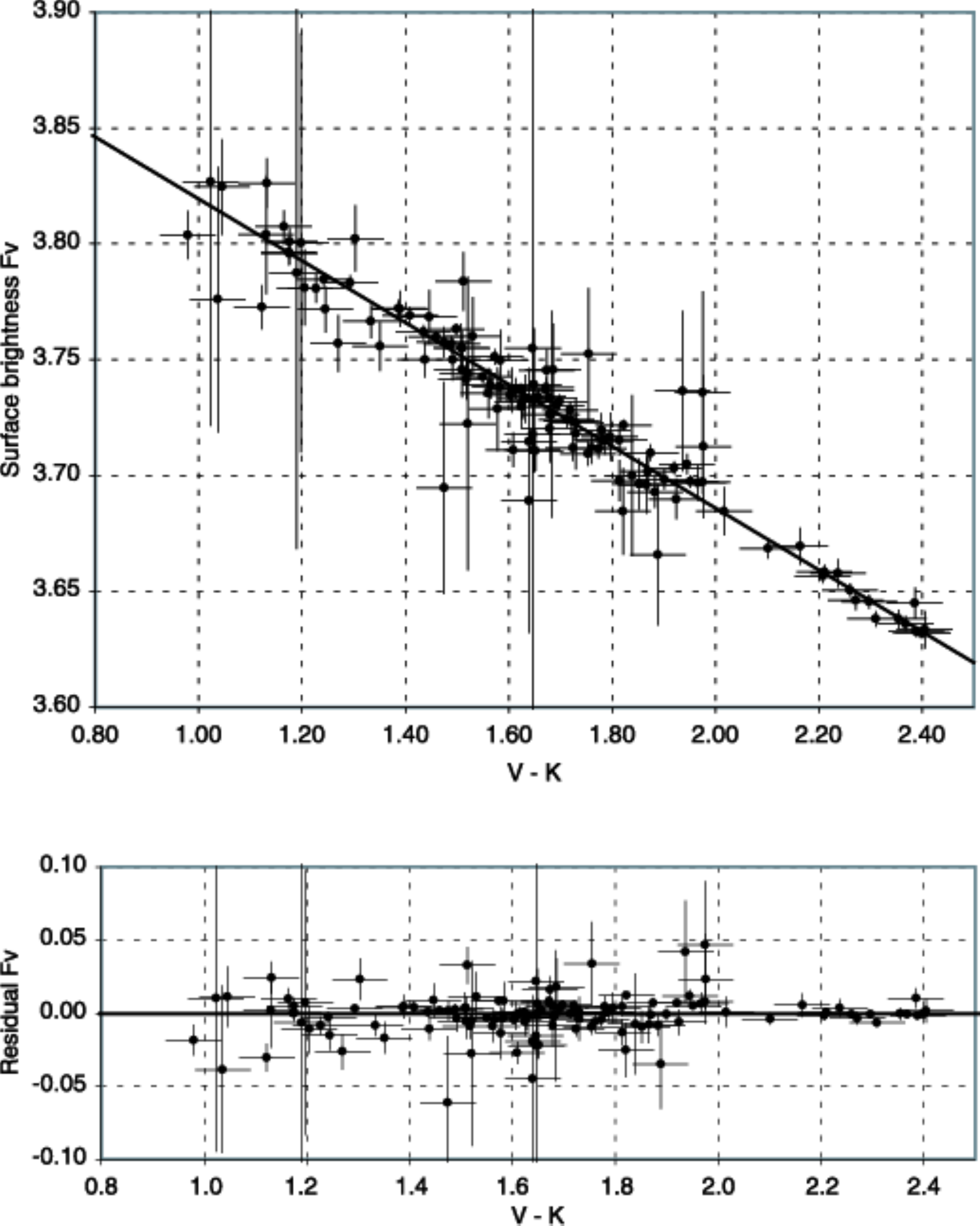}
\caption{Linear fit of $F_v$--$(V-K)$ (upper part) and the corresponding residuals (lower part). Figure from Kervella \emph{et al.} (2004a). \label{fig2}}
\end{figure}

The issue of calibration of the surface brightness by means of observed Cepheid angular diameters has also  been addressed recently. The first demonstration of a surface brightness relation using observed angular diameters was by Nordgren \emph{et al.} (2002) using data from several interferometers.  Based on 59 angular diameter measures of three Cepheids, they found an $F_v$--$(V-K)$ relation consistent with that found for non-variable stars by Fouqu\'e and Gieren (1997).  Kervella \emph{et al.} (2004a) resolved the angular pulsation curves of seven Cepheids with the VLT interferometer.  Figure 2 shows the combined infrared surface brightness relation for these stars. They demonstrated that the observed  surface brightness relation with the smallest scatter is the $V,(V-K)$ relation, that the slope of the surface brightness relation is independent of the period of the Cepheid to within their uncertainties, that the  previously determined surface brightness relation of Fouqu\'e and Gieren (1997), based on angular diameters of non-variable stars, matches very closely the one in Figure 2 based on Cepheid angular diameters, and that the surface brightness relations of Cepheids and main sequence stars of similar $(V-K)$ are essentially identical.   It is not possible to overstate the importance  of these papers to establishing the validity of surface brightness distance measures. 

As to the precision of the infrared surface brightness method, the Bayesian calculations published by Barnes \emph{et al.} (2005) for 38 Galactic Cepheids had a typical random uncertainty in measured distance of $\pm$4\%.  Gieren \emph{et al.} (2005) obtained a similar uncertainty for LMC Cepheids that had periods similar to the Galactic ones studied by Barnes  \emph{et al.} The Gieren \emph{et al.} study demonstrates that the surface brightness method makes it possible to determine with confidence quasi-geometric distances to Cepheid variables at any distance for which the requisite photometry and radial velocities can be obtained.  

\subsection{Interferometric Pulsation Distances}
It is obvious that a superior, if limited, application of the surface brightness method for determining Cepheid distances would use direct measurement of the angular diameters.  Interferometric pulsation distances are superior in that they do not require inference of the angular diameter from a color index and because they are fully independent of reddening.  They are limited because only a modest number of Cepheids are close enough to permit measurement of the angular diameter throughout the pulsation cycle.   Thus there are few such measures and some of those have rather large uncertainties.
\begin{table}
\begin{tabular}{lrrrrrl}
\hline
  \tablehead{1}{l}{b}{Star\\}
  & \tablehead{1}{r}{b}{Log P\\(days)}
  & \tablehead{1}{r}{b}{$\pi$\\(mas)}
  & \tablehead{1}{r}{b}{$\sigma$($\pi$)\\(mas)} 
  & \tablehead{1}{r}{b}{Distance\\(pc)} 
  & \tablehead{1}{r}{b}{$\sigma$($d$)\\($\%$)}
  & \tablehead{1}{l}{b}{Source\\} \\
\hline
$\delta$ Cep & 0.72 & 3.52 & 0.10 & 284 & 2.8 & M\'erand \emph{et al.} (2005)\\
Y Sgr & 0.76 & 1.96 & 0.62 & 510 & 31.6 & M\'erand \emph{et al.} (2009)\\
$\eta$ Aql & 0.85 & 3.31 & 0.05 & 302 & 1.5 & Lane \emph{et al.} (2002)\\
W Sgr & 0.88 & 2.76 & 1.23 & 362 & 44.6 & Kervella \emph{et al.} (2004c)\\
$\beta$ Dor & 0.99 & 3.05 & 0.98 & 328 & 3.1 & Kervella \emph{et al.} (2004c), Davis \emph{et al.} (2006)\\
$\zeta$ Gem & 1.01 & 2.91 & 0.31 & 344 & 10.6 & Lane \emph{et al.} (2002)\\
Y Oph & 1.23 & 2.16 & 0.08 & 463 & 3.7 & M\'erand \emph{et al.} (2007)\\
$l$ Car & 1.55 & 1.90 & 0.07 & 526 & 3.7 & Kervella \emph{et al.} (2004b),Davis \emph{et al.} (2009)\\
\hline
\end{tabular}
\caption{Cepheids with interferometric pulsation parallaxes.  Adapted from Fouqu\'e \emph{et al.} 2007.}
\label{tab:1}
\end{table}
The observational process of measuring stellar angular diameters is well discussed in the literature, $e. g.$, Lane \emph{et al.} (2002),  and inappropriate to include here. Nonetheless, it is important to recognize that the reduction of observed fringe visibilities to angular diameters requires a model for the light distribution in the source.  The standard assumption is that of a uniform intensity disk, hence the "uniform-disk angular diameter" that is usually quoted.  By means of a theoretically established limb-darkening curve, the uniform-disk diameter may be converted to the (larger) limb-darkened diameter.   This, together with the need for a velocity $p$-factor, is why I denote interferometrically determined Cepheid distances as  quasi-geometric.

The first measurement of an interferometric angular diameter of a Cepheid was by Mourard \emph{et al.} (1997) for $\delta$ Cep, followed soon thereafter by Nordgren \emph{et al.} (1999, 2000)  (Polaris $\eta$ Aql, $\delta$ Cep and $\zeta$ Gem),  Lane \emph{et al.} (2000) ($\zeta$ Gem) and others.  While some of these works showed evidence of resolved pulsation, the uncertainties were large enough to prohibit precise distance determination.  That situation was soon improved.  Figure 3 shows the angular pulsation curve for $l$ Car by Davis \emph{et al.} (2009) demonstrating the quality of recent measurements.  Davis \emph{et al.}  determined a distance to $l$ Car with random uncertainty of $\pm$3\%. 

\begin{figure}
 \includegraphics[height=.2\textheight]{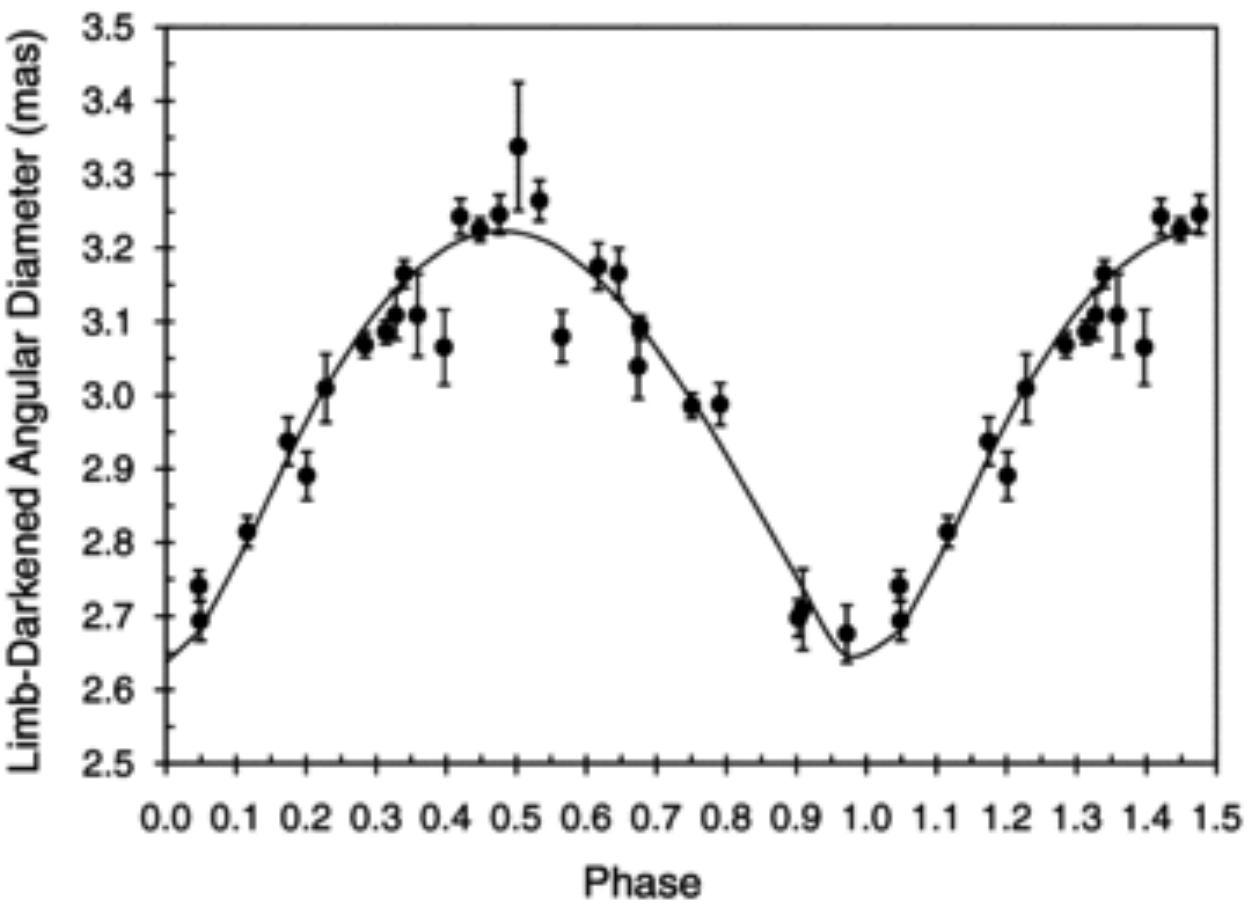}
\caption{Observed angular diameters (points) of $l$ Car compared to scaled linear displacements (smooth curve). Data from SUSI. Figure from Davis \emph{et al.} (2009). \label{fig3}}
\end{figure}

Currently there are eight Cepheids with distances measured through interferometric pulsation parallaxes.  They are listed in Table 5 of Fouqu\'e \emph{et al.} (2007) which is reproduced here as Table 1. I have added to Table 1 the distances and percentage uncertainty in the distance and have reordered the list by period.  The  mean percentage uncertainty in distance for the eight Cepheids is 18\%, but this is dominated by three stars with large uncertainties.  If those are discarded, the mean is 7.2\%.   While interferometric pulsation distances show promise of being superior in precision to infrared surface brightness distances, that promise is not yet fulfilled.  

Recently there has been evidence presented that some of the Cepheids observed for interferometric pulsation distances show evidence of circumstellar material that may affect the interferometry (Kervella \emph{et al.} 2006, M\'erand \emph{et al.} 2006). M\'erand \emph{et al.} estimate the effect on the angular diameter of $\delta$ Cep at about $1\%$.  However, see the paper by N. Evans (2009, these proceedings) in which she did not detect expected circumstellar material.  While a worrisome observation, we will leave it for future investigation and assume here that the interferometric pulsation distances are not affected significantly.  

Moskalik and Gorynya (2005) identify seven additional Cepheids for which resolution of the pulsation cycle may be observed with currently operating interferometers.  These are TT Aql, U Car, X Cyg, T Mon, RS Pup, RZ Vel, and SV Vul.  

\subsection{Trigonometric Distances}
In the preceding sections we have seen that Cepheid distances can be determined with random uncertainties of a few percent by means of surface brightness pulsation distances and interferometric pulsation distances.  The first trigonometric parallaxes to approach this standard for Cepheids were obtained by the \emph{Hipparcos} mission (Perryman \emph{et al.} 1997).   \emph{Hipparcos} measured the parallaxes of numerous Cepheids but very few were individually useful.   

The utility of trigonometric parallaxes for Cepheids took a major leap forward when Benedict and collaborators used one of the Hubble Space Telescope fine guidance sensors (FGS1a) to measure, first the parallax of $\delta$ Cep, and later nine additional Cepheid parallaxes (Benedict \emph{et al.} 2002, 2007).  The HST fine guidance system is intended for spacecraft control, and turns out to be extraordinarily effective for astrometry.  In the papers  cited above, Benedict \emph{et al.} describe in detail how the relative parallaxes are measured within the FGS1a field of view, and how those relative parallaxes are converted to absolute parallaxes through ground-based observations of the reference stars. 

Table 2 shows the HST parallaxes for ten Cepheids as determined by Benedict 's team.  The mean uncertainty  in distance is 8.0\%, which is comparable to the mean uncertainty of the five best interferometric pulsation distances (7.2\%).  

\begin{table}
\begin{tabular}{lrrrrr}
\hline
  \tablehead{1}{l}{b}{Star\\}
  & \tablehead{1}{r}{b}{Log P\\(days)}
  & \tablehead{1}{r}{b}{$\pi$\\(mas)}
  & \tablehead{1}{r}{b}{$\sigma$($\pi$)\\(mas)} 
  & \tablehead{1}{r}{b}{Distance\\(pc)} 
  & \tablehead{1}{r}{b}{$\sigma$($d$)\\($\%$)} \\
\hline
RT Aur & 0.57 & 2.40 & 0.19 & 417 & 7.9 \\
T Vul & 0.65 & 1.90 & 0.23 & 526 & 12.1 \\
FF Aql & 0.65 & 2.81 & 0.18 & 356 & 6.4 \\
$\delta$ Cep & 0.73 & 3.66 & 0.15 & 273 & 4.0 \\
Y Sgr & 0.76 & 2.13 &0.29 & 469 & 13.6 \\
X Sgr & 0.85 & 3.00 & 0.18 & 333 & 6.0 \\
W Sgr & 0.88 & 2.28 & 0.20 & 438 & 8.8 \\
$\beta$ Dor & 0.99 & 3.14 & 0.16 & 318 & 5.1 \\
$\zeta$ Gem & 1.01 & 2.78 & 0.18 & 360 & 6.5 \\
$l$ Car & 1.55 & 2.01 & 0.20 & 497 & 9.9 \\
\hline
\end{tabular}
\caption{Cepheids with trigonometric parallaxes from Benedict \emph{et al.} 2007.}
\label{tab:2}
\end{table}

Recently van Leeuwen \emph{et al.} (2007) used the revised calibration of \emph{Hipparcos} parallaxes to investigate the Leavitt Relation.  It is useful to compare the \emph{Hipparcos}  parallax with the ones from HST.  Figure 4 compares the original \emph{Hipparcos} parallaxes to those from HST, and Figure 5 compares the revised \emph{Hipparcos}  parallaxes.  The mean difference for the original  \emph{Hipparcos} parallaxes is $-0.229\pm0.534$ mas and for the revised parallaxes, $-0.005\pm1.093$ mas.  It is remarkable that the revised  \emph{Hipparcos} and HST parallaxes agree almost perfectly in the mean but the scatter is much larger than was the case for the original  \emph{Hipparcos} parallaxes.  This is largely a result of two outliers (RT Aur and Y Sgr).   If those two stars are removed, the new \emph{Hipparcos} and the HST parallaxes have this mean difference: $+0.123\pm0.405$ mas.  This is a very modest  improvement in consistency over the original  \emph{Hipparcos} catalog.

\begin{figure}
 \includegraphics[height=.3\textheight]{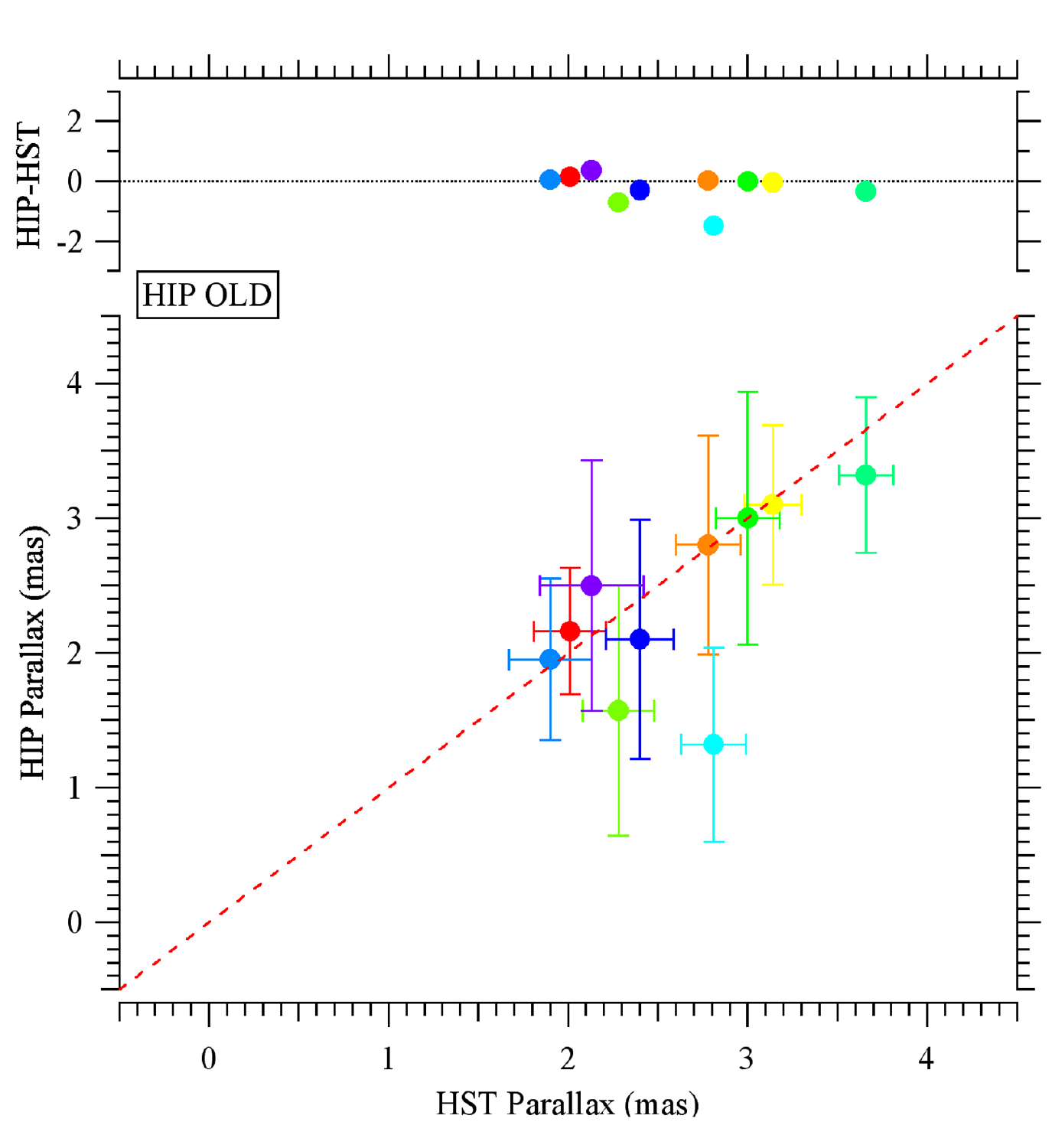}
\caption{Comparison of the original \emph{Hipparcos} parallaxes for Cepheids with those from HST. Figure courtesy of Fritz Benedict. \label{fig4}}
\end{figure}

\begin{figure}
 \includegraphics[height=.3\textheight]{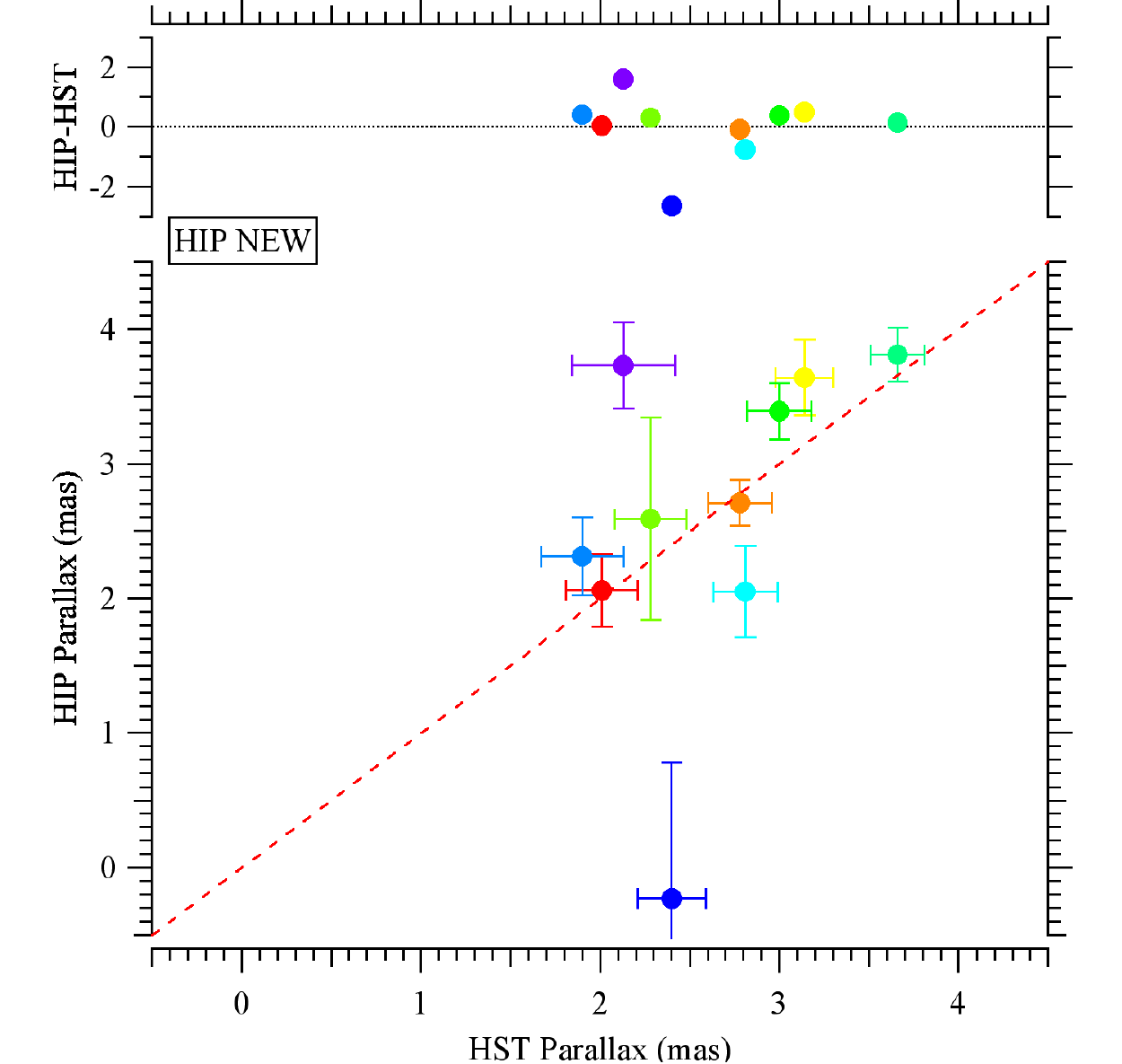}
\caption{Comparison of the revised \emph{Hipparcos} parallaxes for Cepheids with those from HST. Figure courtesy of Fritz Benedict. \label{fig5}}
\end{figure}

\begin{figure}
 \includegraphics[height=.3\textheight]{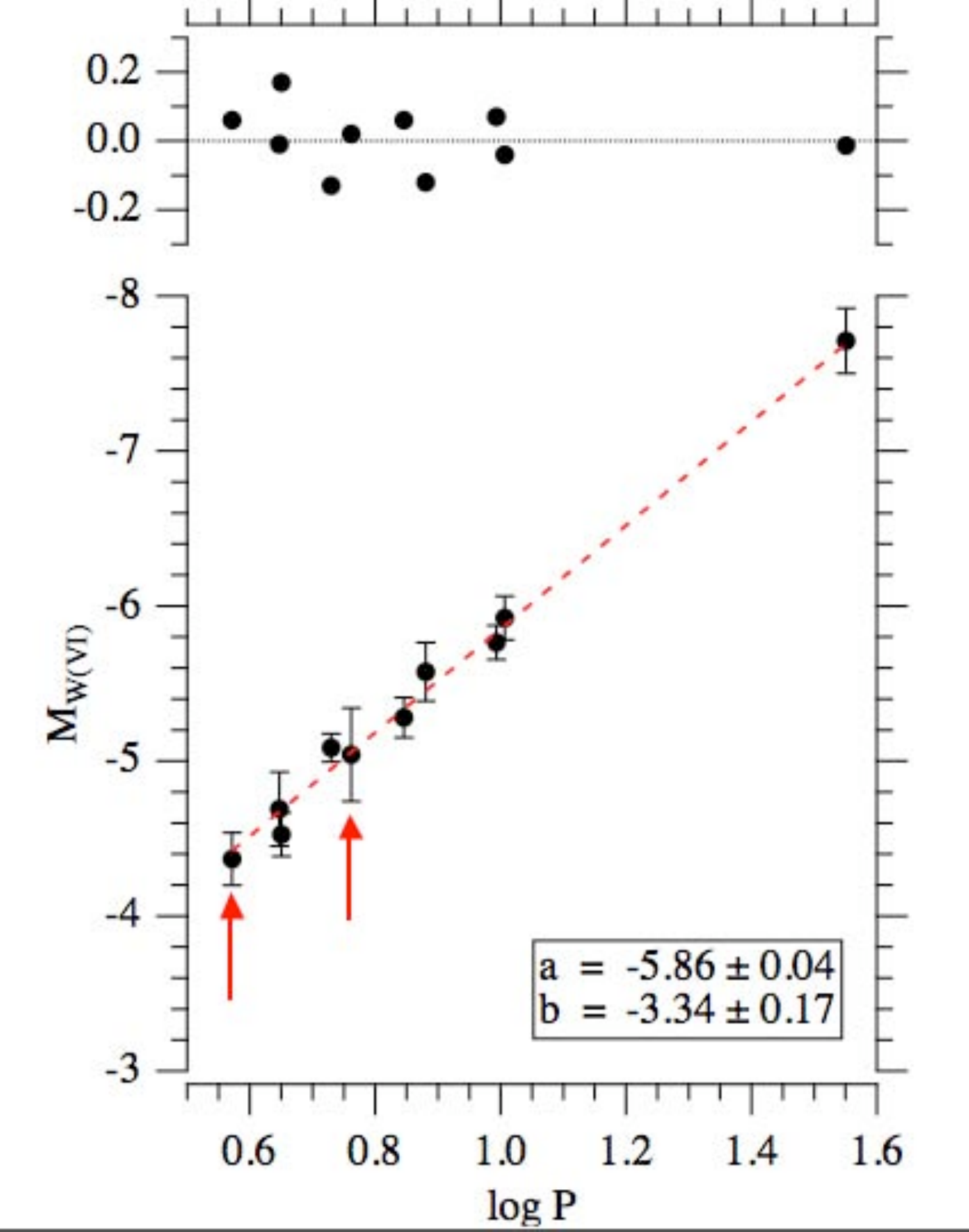}
\caption{Wesenheit Leavitt Law using HST parallaxes. Arrows indicate the stars RT Aur and Y Sgr. Figure courtesy of Fritz Benedict. \label{fig6}}
\end{figure}

\begin{figure}
 \includegraphics[height=.3\textheight]{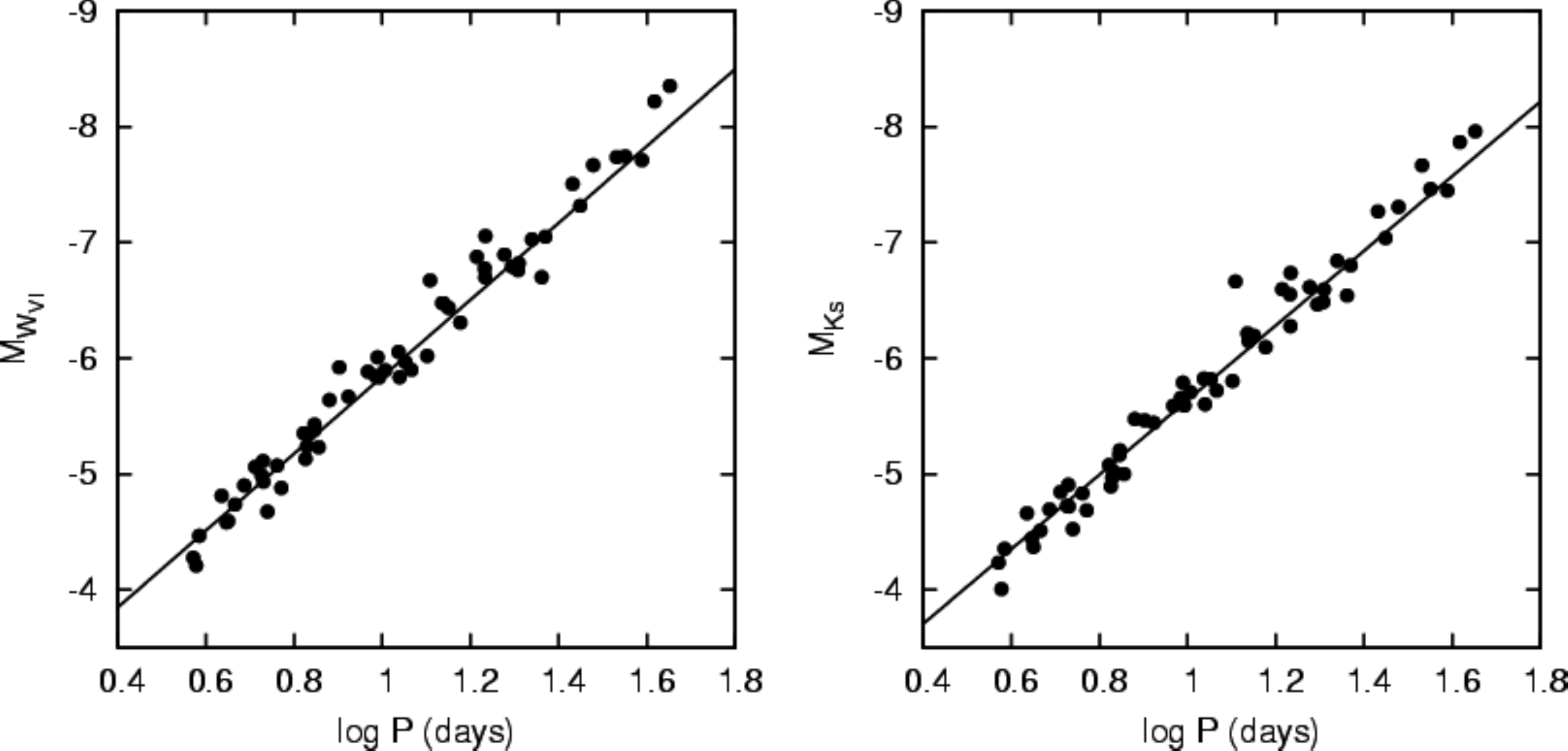}
\caption{A geometrically determined Leavitt Law in the Wesenheit magnitude and the $K$ magnitude.  The solid line is the OGLE slope for the LMC. Figure from  Fouqu\'e \emph{et al.} 2007. \label{fig7}}
\end{figure}

Because the two outliers in Figure 5 suggest that either the \emph{Hipparcos} results or the HST results are subject to unexpectedly large errors, it is important examine the quoted uncertainties in somewhat more detail.   Table 1 of van Leeuwen \emph{et al.} (2007) gives the uncertainties of the \emph{Hipparcos} Cepheid parallaxes in common with HST parallaxes.  Setting aside RT Aur and Y Sgr for the moment, the mean quoted uncertainty of an \emph{Hipparcos} Cepheid parallax is $\pm0.31$ mas, compared to $\pm0.20$ mas for the HST Cepheid parallaxes.  Adding these values in quadrature suggests that the scatter in Figure 5 (again, ignoring the outliers) should be $\pm0.363$ mas, which compares well with the actual  $\pm0.405$ mas and suggests that the quoted uncertainties are realistic.  

On the other hand, if we include the two outliers, the mean quoted  \emph{Hipparcos} uncertainty is $\pm0.38$ mas and the HST, $\pm0.20$ mas.  When added in quadrature, these yield $\pm0.429$ mas, which is far from the actual scatter of $\pm1.093$ mas.  Which set of results causes the outliers?

In Figure 6 I show a Wesenheit Leavitt Law based on the HST parallaxes.  The arrows denote the locations of RT Aur and Y Sgr.  It is clear that they lie well within the scatter band of the HST relation. Had I used the \emph{Hipparcos}  parallaxes for the outliers, RT Aur would lie at least 2.4 mag above the relation ('at least' because its  parallax is negative) and Y Sgr would lie 1.2 mag below it.   I conclude that the anomalies lie within the \emph{Hipparcos} data set.   RT Aur lies 2.6 \emph{Hipparcos} $\sigma$ from the HST result and Y Sgr lies 5.0 $\sigma$ away, which in a sample of ten stars suggests that some of the quoted  \emph{Hipparcos} uncertainties are not Gaussian.  

As in the two previous sections I close this one with a statement about the likelihood of adding more Cepheid distances by this technique in the near future.  Each of the Cepheid parallaxes in Table 2 required eleven orbits, appropriately timed.  The three Cepheids beyond 450 pc have a mean uncertainty of $\pm12\%$.  Significantly more orbits per Cepheid would be needed to obtain parallaxes of Cepheids beyond  500 pc and to higher precision.  To increase the sample significantly in the very near future would require considerable support within the HST TAC.   More likely we will have to await the space missions GAIA and SIM.

\subsection{Fundamental Leavitt Laws} 

\begin{table}
\begin{tabular}{lcccr}
\hline
  \tablehead{1}{l}{b}{Band}
  & \tablehead{1}{c}{b}{Slope}
  & \tablehead{1}{c}{b}{Intercept}
  & \tablehead{1}{c}{b}{$\sigma$} 
  & \tablehead{1}{r}{b}{N}  \\
\hline
$B$ & $-2.289\pm0.091$ & $-0.936\pm0.027$ & 0.207 & 58 \\ 
$V$ & $-2.678 \pm 0.076$ & $-1.275 \pm 0.023$ & 0.173 & 58 \\ 
$R_c$ & $-2.874 \pm 0.084$ & $-1.531 \pm 0.025$ & 0.180 & 54 \\ 
$I_c$ & $-2.980 \pm 0.074$ & $-1.726 \pm 0.022$ & 0.168 & 59 \\ 
$J$ & $-3.194 \pm 0.068$ & $-2.064 \pm 0.020$ & 0.155 & 59 \\ 
$H$ & $-3.328 \pm 0.064$ & $-2.215 \pm 0.019$ & 0.146 & 56 \\ 
$K_s$ & $-3.365 \pm 0.063$ & $-2.282 \pm 0.019$ & 0.144 & 58 \\ 
$W_{vi}$ & $-3.477 \pm 0.074$ & $-2.414 \pm 0.022$ & 0.168 & 58 \\ 
$W_{bi}$ & $-3.600 \pm 0.079$ & $-2.401 \pm 0.023$ & 0.178 & 58  \\ 
 \hline 
 \end{tabular} 
 \caption{Galactic Leavitt Laws from fundamental distances.  Table adapted from Fouqu\'e \emph{et al.} 2007.}
 \label{tab:Fouque} 
 \end{table}
 
 The preceding discussion has laid the foundation for a Leavitt Law based on geometric and quasi-geometric distances to fundamental-mode Galactic Cepheids.  There are seventy Cepheids with distances from the infrared surface brightness method, eight from the interferometric pulsation method and ten from HST trigonometric parallaxes.  Fouqu\'e \emph{et al.} (2007) have combined  these results into a single Leavitt Law in a variety of bands (Table 3). (The fourth column gives the scatter of the absolute magnitudes about the fit.)  When there are multiple distances, they adopted a distance based on HST parallaxes, if available, and then chose from interferometric pulsation distance, infrared surface brightness distance, and  \emph{Hipparcos}  distance on the basis of precision.  Because of overlap in distance measures  and variations in the quality of the photometry, the relations contain up to 59 distances.  In Figure 7 we  show their results for the Wesenheit magnitude and the $K_s$ magnitude compared to the OGLE slope for LMC Cepheids (Udalski \emph{et al.} 1999).

It is important to examine whether there are any  systematic differences between these three  distance  indicators.   Fouqu\'e \emph{et al.} (2007) quote that the infrared surface brightness $W_{vi}$ magnitudes differ from the ones determined from HST parallaxes by $0.01\pm0.03$ mag.  I have computed from their results that the interferometric pulsation distances give $W_{vi}$ magnitudes that differ from the HST ones by $0.03\pm0.13$ mag.  Clearly these fundamental methods agree  with each  other.   

\section{The p-factor}
Until now I have not addressed a  potential systematic error common to the infrared surface brightness distances and the interferometric pulsation distances: the velocity $p$-factor.  Distances by both of these methods scale directly with the adopted $p$-factor. The other potential systematic error in interferometric pulsation distances, the limb-darkening correction, will not be discussed here. See Marengo \emph{et al.} (2002, 2003).

The observed radial velocity is not the pulsational velocity of the stellar atmosphere because of geometrical  projection, limb-darkening, choice of measurement technique, choice of lines measured ($i.e.$, line depth), spectral resolution, and still more parameters.  The effect of these parameters is approximated by multiplying the observed radial  velocity by a $p$-factor.  From these parameters, it is clear that $p$ will vary from Cepheid to Cepheid and with pulsation phase for an individual Cepheid.  Moreover, both the infrared surface brightness method and the interferometric pulsation method require that the $p$-factor link the radial velocity curve to the pulsation curve at the level of the atmosphere that yields the photometry or visibility function.  Fouqu\'e \emph{et al.} (2007) give a brief summary of historical papers on the $p$-factor.  

For the past two decades the $p$-factor chosen for surface brightness calculations has usually been that given by eq. (8) of Gieren \emph{et al.} (1989):

\begin{equation}
p=1.39 - 0.03logP \label{p-factor}
\end{equation}

\noindent
where $P$ is the period of pulsation in days.  They developed this relation as a simplified fit to the theoretical  calculations of $p$ by Hindsley and Bell (1986).   It accounts for the change in $p$ in their models due to mean effective temperature and surface gravity of the Cepheid (using period as a proxy) and ignores any variation with pulsation phase or other factors.  Hindsley and Bell's calculations were appropriate to radial  velocities determined  by cross-correlation radial velocity meters. 

In recent years the success of the infrared surface brightness method and the interferometric pulsation   method generated much interest in establishing the appropriate $p$-factor.  These  efforts were both observational and theoretical. 

The first observational determination of a $p$-factor for a Cepheid was by M\'erand \emph{et al.} (2005).  They used the HST parallax of   $\delta$ Cep to invert the interferometric pulsation method.  Given the distance and the observed angular diameter variation, they determined that $p$-factor that would match the scaled displacement curve to the angular diameters.  Their value  is $p=1.27\pm0.06$. Groenewegen (2007) later extended the analysis to seven stars using the additional  HST parallaxes publish by Benedict \emph{et al.} (2007).  He found $p=1.27\pm0.05$.  He did not find any period dependence.  

The second observational determination was by Gieren \emph{et al.} (2005) using their infrared surface brightness distances to Galactic and LMC Cepheids.  When Gieren \emph{et al.} used eq. \eqref{p-factor} in determination of LMC Cepheid distances, they found a strong dependence of the distance modulus upon period -- an unphysical result. After ruling out other possibilities, they concluded that the $p$-factor eq. \eqref{p-factor}  was incorrect and determined a new one that (1) yielded a zero slope in the period-distance plane (affects the slope in eq. \eqref{p-factor}),  and (2) yielded agreement within the Galaxy in Cepheid $W_{vi}$ magnitudes between infrared surface brightness distances and ZAMS fitting distances (affects the zero  point in eq. \eqref{p-factor}).  Their result, as modified by Gieren \emph{et al.} (2009, these proceedings) is 

\begin{equation}
p=1.52(\pm0.02) - 0.17(\pm0.03)logP \label{LMC}
\end{equation}

Finally, Benedict \emph{et al.} (2007) determined $p$ for T Vul by  inverting the $(V-R)$ surface brightness method using their new HST parallax.  The result, while rather uncertain, $p=1.19\pm0.16$, is another independent determination.

Using very high resolution spectra, Nardetto \emph{et al.} (2004, 2006, 2007, 2009a, 2009b) have examined Cepheid atmospheres and their motions extensively, including a calculation of the $p$-factor. When based on observations of the single line Fe I \emph{$\lambda$4896}, they determined a dependence of $p$ upon period to be (Nardetto \emph{et al.} 2007)

\begin{equation}
p=1.376(\pm0.023) - 0.064(\pm0.02)logP \label{Nardetto1}
\end{equation}

\noindent
whereas observations based on velocities determined by cross-correlation gave (Nardetto \emph{et al.} 2009a) gave 

\begin{equation}
p=1.31(\pm0.06) - 0.08(\pm0.05)logP \label{Nardetto2}
\end{equation}

\noindent
They note that the cross-correlation method of measuring velocities overestimates the velocity curve amplitude and thus the correction factor to pulsational  velocities eq. \eqref{Nardetto2} must be smaller.  This indicates that researchers must ensure that the $p$-factor used in their infrared surface brightness and interferometric pulsation calculations be the correct one for the velocities adopted. 

It is interesting to put these results into context.  In Figure 8 I show the $p$-factor appropriate to the period of $\delta$ Cep according to each of the above studies.  Several conclusions may be drawn from the figure.  First, the value approximated from the work of Hindsley and Bell (1986), source (1), is too large compared to recent determinations in the Galaxy.  A too large $p$-factor yields too large distances in the  surface brightness and interferometric distance methods.  Second, the observational results in the Galaxy and the theoretical results for cross-correlation velocities are consistent (sources 2,4,5,7).  This is very encouraging as the observational work is based on cross-correlation velocities.  Third, the result by Gieren \emph{et al.} (2005, 2009), source 3, using LMC Cepheids is larger  by somewhat more than one sigma than the purely  Galactic determinations.  As the chain or reasoning in Gieren \emph{et al.} (2005) seems strong, one immediately suspects a metallicity effect.  Finally, the $p$-factor for the single line Fe I \emph{$\lambda$4896}, source (6), is larger than the other Galactic determinations, as expected from Nardetto \emph{et al.} (2009a) 

\begin{figure}
 \includegraphics[height=.3\textheight]{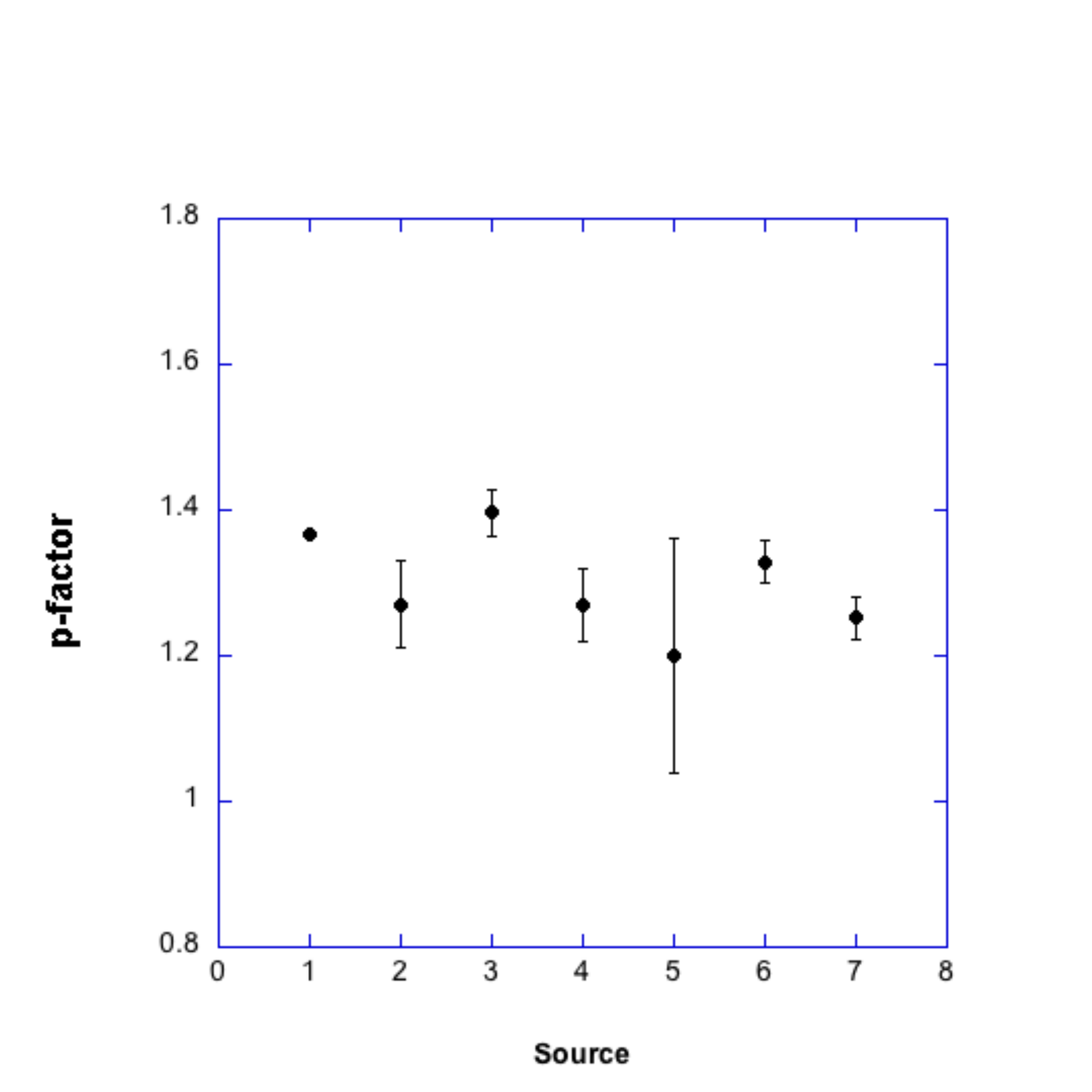}
\caption{$p$ values inferred for $\delta$ Cep from each of the recent $p$-factor studies. Sources: 1) Gieren \emph{et al.} 1989; 2) M\'erand  \emph{et al.} 2005; 3) Gieren \emph{et al.} 2005, 2009; 4) Groenewegen 2007; 5) Benedict \emph{et al.} 2007; 6) Nardetto  \emph{et al.} 2007; 7) Nardetto  \emph{et al.} 2009a.   \label{fig8}}
\end{figure}

Figure 8 shows only a snapshot of $p$ at the period of  $\delta$ Cep.  The dependence upon pulsation period has only one observational determination, in the LMC (Gieren \emph{et al.} 2005, 2009), and one recent theoretical determination, appropriate to Galactic metallicity (Nardetto  \emph{et al.} 2007, 2009a). These differ by a factor of two in slope of  the $p$-factor with period.  In order to improve this situation observationally, we need  more Cepheid parallaxes of high quality to use with more resolved angular pulsation curves in an inverted interferometric  pulsation calculation. It may be some time before this is possible.   Improvement theoretically would be to determine if the difference between the observed LMC slope and the Galactic slope can be understood as a result of differences in the Cepheid atmospheres or whether it requires some other explanation.  

In the previous section on  Leavitt Laws we saw that the three fundamental methods of Cepheid distance determination agree with each other to better than 2\% in distance.  The distance determinations that led to this agreement in Fouqu\'e \emph{et al.} (2007) used a dependence of $p$ upon period quite close to that in eq. \eqref{Nardetto1}.  That relation is appropriate to velocity measures using the line Fe I \emph{$\lambda$4896}, not cross-correlation velocities, which is what the data they had would require.  (This is not a criticism of Fouqu\'e \emph{et al.} as the work by Nardetto \emph{et al.} 2009a on the cross-correlation $p$-factor was still in the future.) The effect of using eq. \eqref{Nardetto1} rather than eq. \eqref{Nardetto2} at the period  of $\delta$ Cep is to increase the distance by about $6\pm4\%$, or about $0.12\pm0.08$ mag in $W_{vi}$.  (There is no difference between the relations in the period dependence of $p$.) The actual difference between the $p$-dependent distances and the HST parallaxes  was  less than 0.03 mag.  The pessimist will  say that this implies a failure of our understanding of $p$; the optimist will say that it shows, to one sigma, that we actually  understand the value of $p$. 

\section{Conclusion}
In this review I have endeavored to summarize recent work  on our understanding of the distance scale of fundamental-mode Cepheids as based on geometric and quasi-geometric distance determinations.  We have seen that the infrared  surface brightness method has reached a level of maturity that permits it to be used as a reliable method for measurement of Cepheid distances in our Galaxy.  The interferometric pulsation method has demonstrated its usefulness as a check on the infrared surface  brightness distance scale and as a method to determine the $p$-factor observationally.  The distances determined by both these methods are found to be in excellent agreement with the high quality parallaxes from HST.

The above agreement between $p$-dependent (quasi-geometric) and trigonometric (geometric) distances notwithstanding, there is still uncertainty in our understanding of the correction factor from  observed radial velocities to pulsational velocities.  The relationship between $p$ and  pulsation period appears  to be different in the Galaxy from that in the LMC.  Until this is resolved, we must be cautious in applying the infrared surface brightness method in environments that differ greatly from the Galaxy.  Within the Galaxy, observational and theoretical determinations of $p$ seem to be in agreement, and this agreement is supported by the agreement between quasi-geometric and geometric distances to Cepheids.  

I am confident that Henrietta Leavitt would be both amazed and pleased at the  progress made  in the past century.


\begin{theacknowledgments}
I express my thanks to G. Fritz Benedict for permission to use figures prepared by him.  I also thank Paul Bradley for assistance with LaTeX.  My interest in the Cepheid distance scale was sparked by J. D. Fernie, reinvigorated by D. S. Evans, sustained over many years by T. J. Moffett and W. P. Gieren, and enhanced by collaborations with many colleagues and students.  For their support I am sincerely grateful.  This material is based in part upon work by TGB while serving at the National Science Foundation. Any opinions, findings, and conclusions or recommendations expressed in this material are those of the author and do not necessarily reflect the views of the National Science Foundation.
\end{theacknowledgments}

\end{document}